# Evidence for topological proximity effect in graphene coupled to topological insulator

Liang Zhang[1], Ben-Chuan Lin[1], Yan-Fei Wu[1], Jun Xu[2], Dapeng Yu[1,3] and Zhi-Min Liao[1,4]*

[1]State Key Laboratory for Mesoscopic Physics, School of Physics, Peking University, Beijing 100871, China

[2]Electron Microscopy Laboratory, School of Physics, Peking University, Beijing 100871, China

[3]Department of Physics, South University of Science and Technology of China, Shenzhen 518055, China

[4]Collaborative Innovation Center of Quantum Matter, Beijing 100871, China

*E-mail: liaozm@pku.edu.cn

**The emergence of topological order in graphene is in great demand for the realization of quantum spin Hall states. Recently, it is theoretically proposed that the spin textures of surface states in topological insulator can be directly transferred to graphene by means of proximity effect. Here we report the observations of the topological proximity effect in the graphene-topological insulator $Bi_2Se_3$ heterojunctions via magnetotransport measurements. The coupling between the $p_z$ orbitals of graphene and the $p$ orbitals of surface states on the $Bi_2Se_3$ bottom surface can be enhanced by applying perpendicular negative magnetic field, resulting in a giant negative magnetoresistance at the Dirac point up to about -91%. An obvious resistivity dip in the transfer curve at the Dirac point is also observed in the hybrid devices, which is consistent with the theoretical predictions of the distorted Dirac bands with unique spin textures inherited from $Bi_2Se_3$ surface states.**

Graphene and surface states of topological insulators (TIs) can be described by two-dimensional (2D) massless Dirac Hamiltonian at the low energy excitations, which can be further modulated by adatom adsorption or interfacing with other functional materials [1-15]. Owing to the high carrier mobility and unique spin textures, TIs and graphene are promising for high speed electronics and spintronics [16]. Recent theories [8-11] have predicted the hybridization of graphene and TIs can create nontrivial spin textures in graphene, even leading to quantum spin Hall states [1]. Generally, the rigorous $\sqrt{3} \times \sqrt{3}$ supercell of graphene stacked with TI is adopted in the calculations [8-11]. For the incommensurate graphene-TI stacking, Zhang *et al.* [10] suggested that the renormalized bands of hybrid graphene still acquire the in-plane spin textures from the surface states of TI even in the presence of surface roughness at the heterointerface. The coupling between graphene and TI has been pursued *via* the angle resolved photoemission spectroscopy (ARPES) experiment [17]. On the other hand, the proximity effect has been demonstrated in both graphene and TI based hybrid devices, such as graphene/$WS_2$ [6,7], graphene/$BiFeO_3$ [18], graphene/EuS [19] and Bi/$TlBiSe_2$ heterostructures [20]. Although the graphene-TI heterostructures have been fabricated to demonstrate the persistent of topological surface state of $Bi_2Se_3$ [21], 2D density of states (DOS) related tunneling process [22, 23] and current-induced spin polarization [24], the proximity effect induced fascinating properties in graphene-TI hybrid devices are still unrevealed. The topological proximity mainly affects the electronic properties of graphene in the vicinity of the Dirac point [8-11]. The strong hybridization will make the energy dispersion nonlinear near the charge neutrality point, resulting in the enhancement of the DOS at the Dirac point in graphene [10, 11]. Here, we report on the anomalous magnetotransport properties at the Dirac point in graphene coupled to $Bi_2Se_3$ nanoribbons.

**Results**

**Gating effect of Graphene-$Bi_2Se_3$ hybrid devices**. High quality $Bi_2Se_3$ nanoribbons grown by the chemical vapor deposition (CVD) method were transferred onto mechanically exfoliated monolayer graphene sheets on 285 nm $SiO_2$/Si substrates (see Methods and Supplementary Fig. 1a-d for details). The schematic diagram of the patterned Hall bar device is shown in Fig. 1a. The two current leads and four voltage probes for Hall measurements are Au/Pd (80 nm/5 nm) electrodes on graphene. The scanning electron microscope (SEM) image of a typical graphene-$Bi_2Se_3$ device is presented in the inset of Fig. 1b (see Supplementary Fig. 1 for more typical devices). The $Bi_2Se_3$ nanoribbon covers the entire transport channel of graphene with the same width. The back gate voltage $V_g$ was used to tune the carrier density of graphene. The band structures of graphene and $Bi_2Se_3$ are illustrated in Fig. 1a. The $Bi_2Se_3$ nanoribbons used here are single crystals and the Fermi level is ~350 meV above the Dirac point of surfaces states as revealed by the ARPES experiment [25, 26]. Such heavy doping in $Bi_2Se_3$ will lead to (i) the enhanced hybridization between graphene and TI, considering that the DOS of TI increases dramatically [10]; (ii) the significant hexagonal warping effect of the surface states [27], which is directly responsible for the out-of-plane spin component up to 12% [28]. In this regard, the out-of-plane spin polarizations in graphene cannot be neglected in the realistic graphene-TI coupling systems, which could be important in the transport. For instance, the out-of-plane spin-orbit torque induced by the Fermi surface warping has been reported in $Bi_2Se_3$-based heterostructures, which is comparable to the in-plane torque below 50 K [29].

Gate-tunable conductivity $\sigma_{xx}$ of the hybrid device is shown in Fig. 1b, reproducing the field effect of graphene. The position of Dirac point has been adjusted to zero volt, *i.e.* $V_g^* \equiv V_g - V_D$, where $V_D$ is the measured Dirac point in $V_g$ and $V_D$ = 25 V for the device presented in Fig. 1b. The approximately linear dependence $\sigma_{xx}(V_g^*) \propto V_g^*$ away from the Dirac point indicates the charged impurity scattering dominated transport [30]. The mobility $\mu = \sigma_{xx}/en$ exceeds 1.5 $m^2V^{-1}s^{-1}$ for both electrons and holes near the Dirac point *via* Hall measurements. Although n-type doped $Bi_2Se_3$ has a large conductivity, the minimum conductivity $\sigma_{xx,min}$ still approaches $4e^2/h$, further excluding the direct electrical contribution from the upper $Bi_2Se_3$. Therefore, we reasonably neglect the influence of the conductivity $\sigma_{TI}$ of $Bi_2Se_3$ bulk and the drag conductivity $\sigma_d$ ($\sigma_d^2 \ll \sigma_{xx}\sigma_{TI}$) near the Dirac point in the following discussions. According to the analytic transport theory for the electron-hole puddles landscape developed by Adam *et al.* [31], the effective carrier density $n^*$ is estimated to be ~ $6.6 \times 10^{10} cm^{-2}$. Besides, the thermally excited carrier density $n_e = n_h \approx 0.52(k_BT/\hbar v_F)^2$ at the charge neutral point, where $k_B$ is the Boltzmann's constant. At T = 1.4 K, $n_e = n_h \approx 1.75 \times 10^6 cm^{-2} \ll n^*$. Therefore, we only need to consider the electron-hole puddles induced by the long-range Coulomb scattering at Dirac point in the low temperature regime.

**Landau levels in graphene-TI hybrid devices.** To demonstrate the unusual properties of graphene-$Bi_2Se_3$ heterointerface, the magnetotransport properties were further systematically measured. Figure 2 shows the resistivity $\rho_{xx}$ and $\rho_{xy}$ as a function of the gate voltage $V_g^*$ under various perpendicular magnetic fields at 1.4 K (see Supplementary Fig. 2 for detailed comparisons). Both positive and negative magnetic fields were applied to the devices. The transfer curves in Fig. 2a,b are shifted linearly proportional to the strength of the applied magnetic field for clarity. Notably, the longitudinal resistivity $\rho_{xx}$ at the Dirac point is largely suppressed and even exhibits a dip under the low negative magnetic field. Although the resistivity peaks of the zeroth Landau level (LL) recover under high negative magnetic fields, the values are still obviously smaller than the values under positive magnetic fields. These observations can be understood in the framework of the graphene-TI proximity effect that was theoretically predicted *via* band calculations in Refs. [8-11]. Under the magnetic field perpendicular to the graphene plane, the unevenly spaced energy spectrum is expressed as $E_N = sgn(N)\sqrt{2e\hbar v_F^2|N|B}$, where $\hbar$ is reduced Planck's constant, the Fermi velocity $v_F$ is $\sim 10^6 \, m/s$, and the LL index N is positive for electrons and negative for holes [12, 13]. The half-filled LLs, such as $N = -3, -2, \pm 1$ can be clearly identified as $\rho_{xx}$ peaks under high magnetic field, and the positions represented by $V_g^{*,p}$ are denoted by red filled circles in Fig. 2a,b. When the Fermi level locates exactly at each LL, the $\rho_{xx}$ peak positions are at $V_g^{*,p} = 4e^2NB/hc_g$, where $c_g$ is the effective capacitance of the back gate. For a fixed Landau index N, the position $V_g^{*,p}$ is linearly dependent on the magnetic field strength $B$, as shown by the dashed lines in Fig. 2a,b. The linear fitting of the experimental data yields the averaged capacitance $c_g \approx 10.7\ nFcm^{-2}$, which is consistent with gate modulation by the SiO$_2$ dielectric layer. Moreover, the slopes of the fitting lines extracted from Fig. 2a,b also follow linear relationship with the Landau index N, as shown in Fig. 2c. The results indicate that the LL structures and Dirac natures are maintained in the hybrid graphene devices under high magnetic field. Although the longitudinal resistivity $\rho_{xx}$ is unusual near the Dirac point under negative magnetic field, the corresponding Hall resistivity $\rho_{xy}$ plateaus in Fig. 2d are precisely quantized at $h/g(N+\frac{1}{2})e^2$ with $g = g_s g_v = 4$. The Landau level-induced quantum oscillations of conductivity away from the Dirac point can also be clearly identified (Supplementary Fig. 3). We should realize that the possible coupling between graphene and TI surface states is the second (or higher) order effect [8-11]. The hopping process between the $p_z$ orbitals of carbon atoms in graphene and $p$ orbitals of the bottom surface states in Bi$_2$Se$_3$ nanoribbons introduces significant influences on the transport properties in graphene near the Dirac point.

**Anomalous magnetotransport features at Dirac point.** Now we discuss the magnetotransport at the Dirac point (or the zeroth LL) in the graphene hybrid device. The evolution of $\rho_{xx}(V_g^*)$ curves near the Dirac point at various temperatures and under negative magnetic fields is shown in Fig. 3a (see Supplementary Fig. 4 for the evolution of $\sigma_{xx}(V_g^*)$). The curves are shifted for clarity. The resistivity at the Dirac point ($\rho_{xx}^D$) versus B is extracted and shown in Fig. 3b,c. In the classical regime, the two-carrier model can be used to describe a zero-gap conductor with the same mobility $\mu$ for electrons and holes, giving $\rho_{xx}^D(B) = \rho_{xx}^D(0)\frac{1+(\mu B)^2}{1+(\beta\mu B)^2}$, where $\beta = \frac{n_e - n_h}{n_e + n_h}$. Thus, at the charge-neutral point ($n_e - n_h \approx 0$), we have $\rho_{xx}^D(B) = \rho_{xx}^D(0)[1 + (\mu B)^2]$ and $\rho_{xy} = 0$. Considering the spatial inhomogeneity of electrons and holes with equal density and mobility, there is a positive magnetoresistivity (MR) $\rho_{xx}^D(B) = \rho_{xx}^D(0)[1 + (\mu B)^2]^{1/2}$ [32, 33]. If there is a small parallel conductivity $\sigma^p$ due to the difference of density and mobility between electrons and holes, the MR is modified as $\rho_{xx}^D(B) = (\sigma_{xx}^D(0)[1 + (\mu B)^2]^{-1/2} + \sigma^p)^{-1}$ [34]. However, the upper classic models cannot explain the MR behaviors in the graphene-Bi$_2$Se$_3$ hybrid devices.

Predicted by theoretical models [8-11], graphene can inherit spin-orbital textures from TI surface states near the Dirac point due to the proximity effect. Accordingly, we summarize the reforming band structures of graphene hybridized with TI surface in Fig. 4. As the interaction between graphene and TI is significant, the fourfold degeneracy of the original graphene bands (Fig. 4a) is partially lifted: (1) the two gapped Rashba-like bands are spin polarized; (2) the other two gapless degenerate bands are antisymmetric combinations of the opposite valleys ($K_0, K_0'$) of the isolated graphene (Fig. 4b). The stronger of the interaction is, the more distorted of the bands is (Fig. 4c). To establish the effective Hamiltonian of the two gapless bands, the

Löwdin perturbation theory to the full Hamiltonian described in Ref. 11 is employed as

$$H_{eff}(k) = (C_2 k^2 + C_3)\hat{z} \cdot (\vec{\sigma} \times \vec{k}) - \frac{C_1}{2}\left((k_x + ik_y)^3 + (k_x - ik_y)^3\right)\sigma_z,$$

where $C_i$ ($i = 1, 2, 3$) is the parameter, $\vec{\sigma} = (\sigma_x, \sigma_y, \sigma_z)$ are the Pauli matrices. Then we can concisely re-write the effective Hamiltonian as $H_{eff} = \sum_{i=x,y,z} d_i(k)\sigma_i$ with momentum-related parameters $d_i(k)$. Moreover, the hexagonal warping effects of TI surface states should be taken into account as the Fermi level resides deeply in the Bi$_2$Se$_3$ bulk conduction band. The warping term can be formulated as [27]

$$H_w^{TIS} = \frac{\lambda}{2}\left((k_x + ik_y)^3 + (k_x - ik_y)^3\right)\sigma_z,$$

where $\lambda$ describes the hexagonal warping strength. When $\lambda = 0$, the formula $H_{eff}$ can roughly describe the gapless bands. In contrast, as $\lambda \neq 0$ (in realistic Bi$_2$Se$_3$ cases), the gapless bands become more non-linear. Note that both $H_{eff}$ and $H_w^{TIS}$ include the out-of-plane spin component $\sigma_z$. Importantly, the interaction of the electron spin $\vec{\sigma}$ with the perpendicular magnetic field $\vec{B}$ ($= B\hat{z}$) can be expressed as $H_Z \propto \vec{\sigma} \cdot \vec{B} = B\sigma_z \propto \sigma_z$. Intuitively, the magnetic field $\vec{B}$ will modify the renormalized bands and influence the topological proximity effect.

Indeed, our main observations (Fig. 3) are consistent with the above analysis. As the underlying graphene is coupled to the bottom surface of Bi$_2$Se$_3$, the direction of the external magnetic field will have distinct influences on the coupling. As shown in Fig. 3b,c, the $\rho_{xx}^D(\vec{B})$ is quite asymmetric between the positive and negative magnetic fields. By reversing the current direction, the mixture of Hall signal origin is excluded for such asymmetry of longitudinal resistivity (Supplementary Fig. 5). From the point of view of topological proximity effect, the asymmetric dependence of longitudinal resistivity $\rho_{xx}$ on magnetic field $B$ indicates that the gapless bands are more non-linear with negative magnetic field and less non-linear with positive field. Apparently, the non-linear bands possess large DOS in the vicinity of the Dirac point (Fig. 4c), leading to the resistivity dip in the transfer curves near the Dirac point and the negative MR of $\rho_{xx}^D$. Moreover, as the renormalized gapped bands are spin-polarized [10, 11], the inherited spin texture in graphene from TI surface states may suppress the scatterings, and then decreases the resistivity. Nevertheless, the hexagonal term could complicate the actual situation. For the negative magnetic field with $|B| \geq 3$ T, the resistivity starts to increase with increasing the magnitude of B, suggesting the saturation of the negative magnetic field induced enhancement of graphene-TI coupling. Under $|B| \geq 3$ T, the inset in Fig. 3b shows that the $\rho_{xx}^D(B > 0)$ and $\rho_{xx}^D(B < 0)$ demonstrate the same dependence with B, where each data point of $\rho_{xx}^D(B < 0)$ has been shifted up with 10 kΩ.

The Zeeman splitting is the most possible origin of the rapid increase of $\rho_{xx}^D$ under $|B| \geq 3$ T at 1.4 K. Each LL splits into two levels with $E_N \pm \Delta E_Z$, where the Zeeman energy $\Delta E_Z = \mu_B B$, and $\mu_B$ is the Bohr magneton. The Fermi distribution function $f(E)$ of the zeroth LL at $E_F = 0$ is $1/[\exp(\frac{\Delta E_Z}{k_B T}) + 1]$. As the factor $\frac{\Delta E_Z}{k_B T}$ is large at low temperatures, $\rho_{xx}^D \propto \exp(\frac{\mu_B B}{k_B T})$ is expected in the Zeeman splitting dominated regime. The experimental data shown in the inset in Fig. 3b are well fitted by $\rho_{xx}^D \propto \exp(\alpha B)$. The parameter $\alpha \approx 0.33$ is obtained from the fitting, which agrees well with the theoretical value $\frac{\mu_B}{k_B T} \approx 0.48$ at 1.4 K. The negative MR can still be

9 / 31

clearly observed at 100 K (Fig. 3c), suggesting the robustness of the proximity effect induced topological transport in graphene. Under $|B| > 3$ T, the linear relationship $\rho_{xx}^D \propto B$ is observed at 100 K (Fig. 3c). Under 14 T, the Zeeman energy gap $2\Delta E_Z$ is corresponding to ~19 K. At high temperatures, the Zeeman splitting is less important and the MR can be described by $\rho_{xx}^D(B) = \rho_{xx}^D(0)[1 + (\mu B)^2]^{1/2}$ for the graphene system at the Dirac point with spatial inhomogeneity of electrons and holes puddles [32, 33]. The linear fitting of the $\rho_{xx}^D$~B under $|B| > 3$ T at temperatures ranged from 20 K to 100 K yields almost the same slope ~ 1.45 kΩ/T (Supplementary Fig. 6), which is ascribed to the temperature insensitive mobility in this temperature region [35]. At the intermediate temperature (1.4 K < T < 19 K), such as 10 K, the magnetic field dependence of $\rho_{xx}^D$ is between exponential form and linear form, as shown in the inset in Fig. 3c.

The negative magnetic field induced enhancement of graphene-TI coupling is further verified by the temperature dependence of resistivity, as shown in Fig. 3d-e. Under zero magnetic field, the $\rho_{xx}^D$ decreases with increasing temperature, which is very different from the pristine monolayer graphene with weak temperature dependence of resistivity [35]. With increasing temperature, the decrease of $\rho_{xx}^D$ can be attributed to the nonlinear energy bands induced by the proximity effect (Fig. 4b,c), because more carriers can be thermally excited to the multiple bands near the charge neutrality point [10, 11]. Interestingly, the hybrid graphene exhibits metallic behaviors, as a small magnetic field such as -1 T is applied. As discussed before, the negative magnetic field can enhance the conducting hybrid states in graphene/Bi$_2$Se$_3$ heterointerface and may suppress the carrier back-scattering in graphene. The reduced thermal perturbation of the spin texture in graphene leads to the decrease of resistance with decreasing temperature. Under B = 14 T, the Zeeman splitting gap gives rise to the rapid rise of $\rho_{xx}^D$ below 20 K, which is consistent with the analysis of the MR under high magnetic fields.

The coupling between graphene and TI also has an influence on the magnetotransport away from the Dirac point in graphene-$Bi_2Se_3$ heterostructures. Distinct from individual graphene, the forward and backward propagating quantum Hall edge states in the hybrid graphene devices can interact with each other *via* the coupled $Bi_2Se_3$ nanoribbon. Specifically, the momentum relaxation *via* the $Bi_2Se_3$ bulk states makes the Hall conductivity deviate from the quantized values at the electron side, as shown in Fig. 5 (see Supplementary Fig. 7 for another example). The scattering between graphene and $Bi_2Se_3$ bulk states is further clearly demonstrated by directly contacting the voltage probes with both graphene and $Bi_2Se_3$ (see Supplementary Fig. 8). It has been theoretically proposed that the topological proximity effect can produce topological edge states in monolayer graphene. The maximum of conductance carried by the quantum spin Hall edge states should be quantized to $2e^2/h$ at the zeroth LL, however, the conductance at B = 3 T observed here at the Dirac point is much larger than $2e^2/h$ (Supplementary Fig. 4). The additional scatterings from the TI bulk states may be responsible for the absence of topological edge state transport here.

**Discussion**

The hybridization between $p_z$ orbitals of graphene and the $p$ orbitals of Bi$_2$Se$_3$ bottom surface states yields a renormalized band structure at the heterointerface. Interestingly, the non-linear gapless bands reside in the bandgap of the Rashba-like spin-polarized bands. We find that the negative magnetic field can tune the coupling strength between graphene and Bi$_2$Se$_3$, leading to a resistivity dip in the vicinity of Dirac point and anomalous magnetoresistivity. As the fourfold degeneracy of graphene is partially lifted (Fig. 4b,c), two Hall conductivity $\sigma_{xy}$ plateaus quantized at $\pm e^2/h$ will be evolving in the N = 0 regime under intermediate magnetic field. Indeed, we have observed such developing $\pm e^2/h$ Hall conductivity plateaus in many graphene-Bi$_2$Se$_3$ samples (Supplementary Fig. 9). Sometimes, the $e^2/h$ Hall conductivity plateau in the electron branch can be blurred by the parallel conducting channels of Bi$_2$Se$_3$, similar with other Hall conductivity plateaus in the electron branch (Fig. 5 and Supplementary Fig. 7). Under 14 T strong magnetic field, the zero Hall conductivity plateau emerges at Dirac point as shown Fig. 5 (also see Supplementary Fig. 7), indicating the gapless bands are finally gapped. The exponential fitting of longitudinal resistivity in Fig. 3b shows that the Zeeman effect is responsible for the evolving gap. Moreover, the renormalized bands due to topological proximity effect will be largely weakened under intense magnetic field, as the time-reversal symmetry can be broken by the magnetic field.

The anomalous magnetotransport results interpreted by topological proximity effect in graphene-Bi$_2$Se$_3$ hybrid suggest another possibility to manipulate the topological properties of quantum matters. Such as, one of the topological states long sought in graphene is quantum spin Hall phase may be realized in topological insulator/graphene/topological insulator sandwiched system, which has been theoretically investigated [8]. Different from the asymmetric graphene/topological insulator configuration [9-11], the wavefunctions in graphene sandwiched by TIs can be hybridized with two TI surfaces. With an intrinsic bulk bandgap, the two-dimensional topological states may emerge in graphene-TI hybrid device.

In summary, we have demonstrated the topological proximity-induced anomalous MR in graphene-TI hybrid devices. The reforming band structures of graphene result in a resistivity dip in the transfer curve near the Dirac point, which can be further enhanced by external negative magnetic field. Our observations should inspire more works to further understand the coupling between graphene and topological insulators, which are valuable to realize exotic topological states based on graphene hybrid devices.

**Methods**

**Device fabrication.** The monolayer graphene was mechanically exfoliated from Kish graphite onto $SiO_2$/Si substrates and identified by Raman spectrum. Then the topological insulator $Bi_2Se_3$ nanoribbon grown by CVD method was directly transferred onto the top of graphene sheet by a micromanipulator. The dry mechanical transfer method can avoid contaminations at the graphene/$Bi_2Se_3$ interface. Standard electron-beam lithography and oxygen plasma etching were employed to shape the underlying graphene to Hall bar. The fabrication processes are schematically illustrated in supplementary Fig. 1. The width of the graphene Hall bar was the same as that of the top $Bi_2Se_3$ nanoribbon. The Au/Pd (80 /5 nm) electrodes were only contacted with graphene, eliminating the direct conduction contribution from the $Bi_2Se_3$ nanoribbon.

**Transport measurements.** The transport measurements were performed in an Oxford cryostat with a variable temperature insert and superconductor magnet. The temperature can be decreased to 1.4 K and the magnetic field can be swept up to 14 T. The electrical signals were measured using a low frequency lock-in technique with the bias current of 0.1 µA.

*Acknowledgement.* This work was supported by National Key Research and Development Program of China (Nos. 2016YFA0300802, 2013CB934600, 2013CB932602) and NSFC (Nos. 11274014, 11234001).

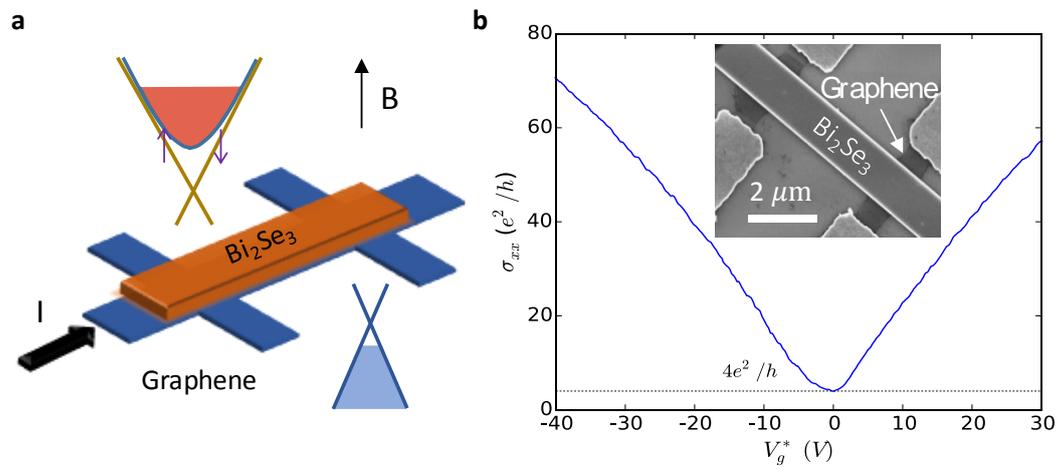

**Figure 1 | Structure of graphene-Bi$_2$Se$_3$ hybrid devices**. (**a**) Schematic diagram of the hybrid device and the band structures of graphene and Bi$_2$Se$_3$. All electrodes are contacted with graphene. (**b**) The gate voltage dependence of conductivity at 1.4 K. The position of the Dirac point has been shifted to zero volt. The inset shows the SEM image of a typical device.

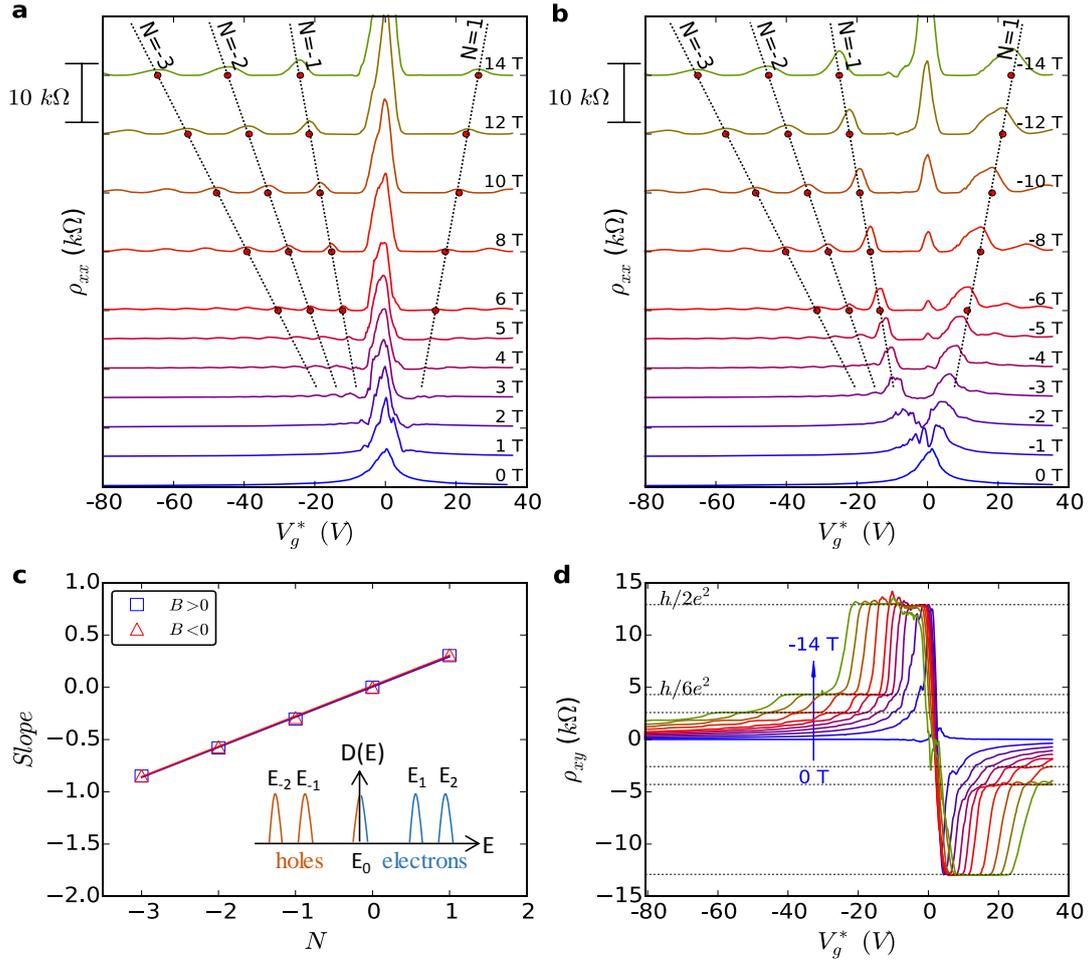

**Figure 2 | Magnetotransport at 1.4 K**. The longitudinal resistivity $\rho_{xx}$ versus gate voltage $V_g^*$ in various (**a**) positive and (**b**) negative magnetic fields. The curves are shifted proportional to the magnetic field strength. The red filled circles indicate the positions $V_g^{*,p}$ of $\rho_{xx}$ peaks. Dashed lines are linear fitting results, demonstrating the $V_g^{*,p} \propto B$. (**c**) The extracted slopes of the dashed lines in Fig. 2a,b versus the Landau index $N$. The data can be linearly fitted, which indicates the LL energy spectrum $E_N \propto \sqrt{|N|B}$. (**d**) The Hall resistivity $\rho_{xy}$ as a function of $V_g^*$ under negative magnetic fields corresponding to Fig. 2b. The dashed lines indicate the quantized resistivity $\pm h/2e^2$, $\pm h/6e^2$, and $\pm h/10e^2$, respectively.

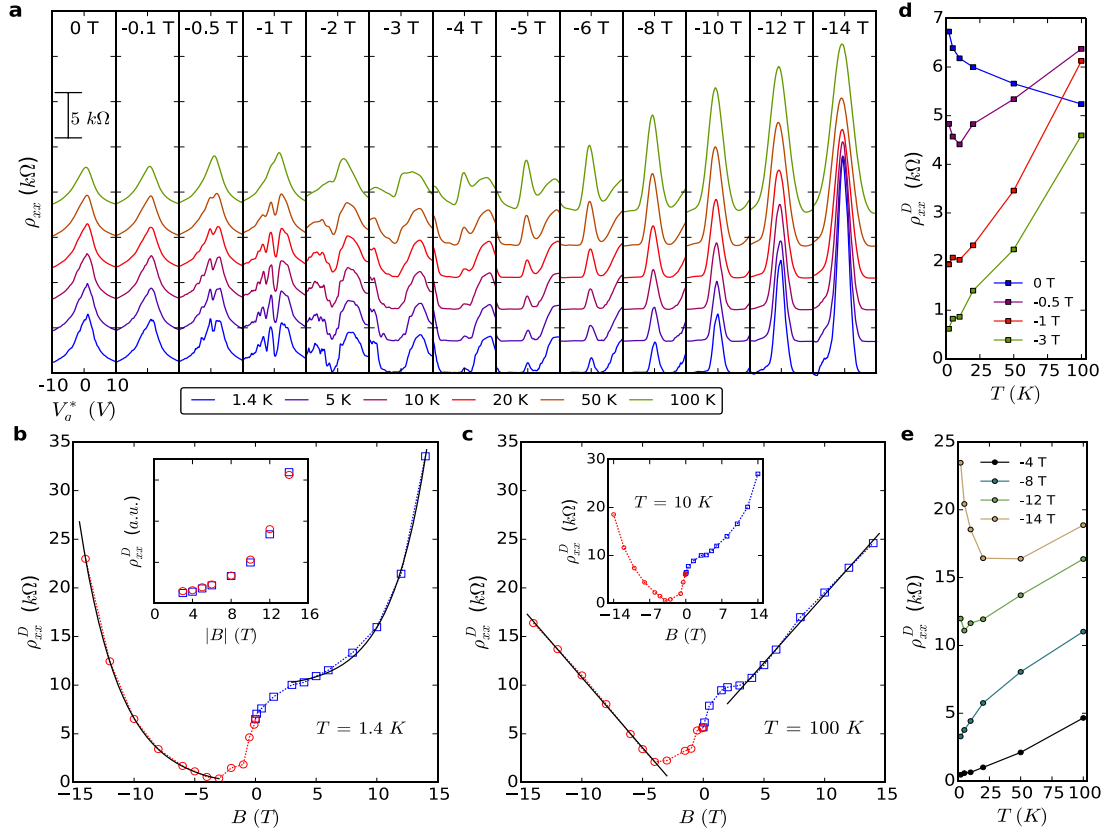

**Figure 3 | Magnetotransport at the Dirac point**. (**a**) The evolution of the $V_g^*$ dependence of $\rho_{xx}$ at various temperatures and negative magnetic fields. Curves are shifted in each panel for clarity. (**b**) The longitudinal resistivity at the Dirac point $\rho_{xx}^D$ versus magnetic field $B$ at 1.4 K. The solid curves show the exponential fitting as $|B| \geq 3$ T. The $\rho_{xx}^D$ vs. $B$ has the same exponential factor for both positive and negative magnetic fields, as demonstrated in the inset, where the $\rho_{xx}^D(B<0)$ data points (open red circles) are shifted up. (**c**) The $\rho_{xx}^D$ vs. $B$ at 100 K. The solid lines show the linear fitting results. The inset shows the $\rho_{xx}^D$ vs. $B$ at 10 K. (**d,e**) The $\rho_{xx}^D$ versus temperature $T$ at various magnetic fields.

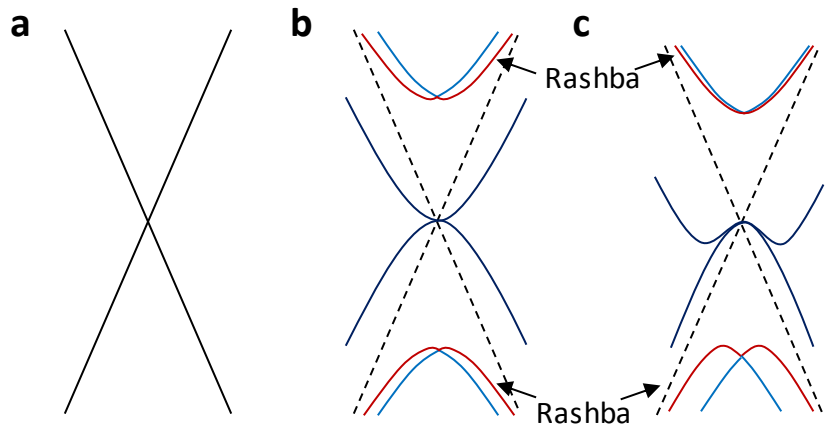

**Figure 4 | Reforming band structures of graphene-topological insulator hybrid structure.** (**a**) Band structure of pristine graphene. (**b,c**) Schematic band structures of graphene hybridized with topological insulator surface. The dashed black lines indicate the original graphene bands with fourfold degeneracy, which is partially lifted in the hybrid structure. The two gapped bands labelled as Rashba are spin polarized, originating from the Rashba spin-orbit coupling. The other two degenerate bands are antisymmetric combinations of the opposite valleys $K_0, K_0'$ of the isolated graphene. The hybrid band structures can be further distorted from **b** to **c** by strengthening the graphene-topological insulator coupling.

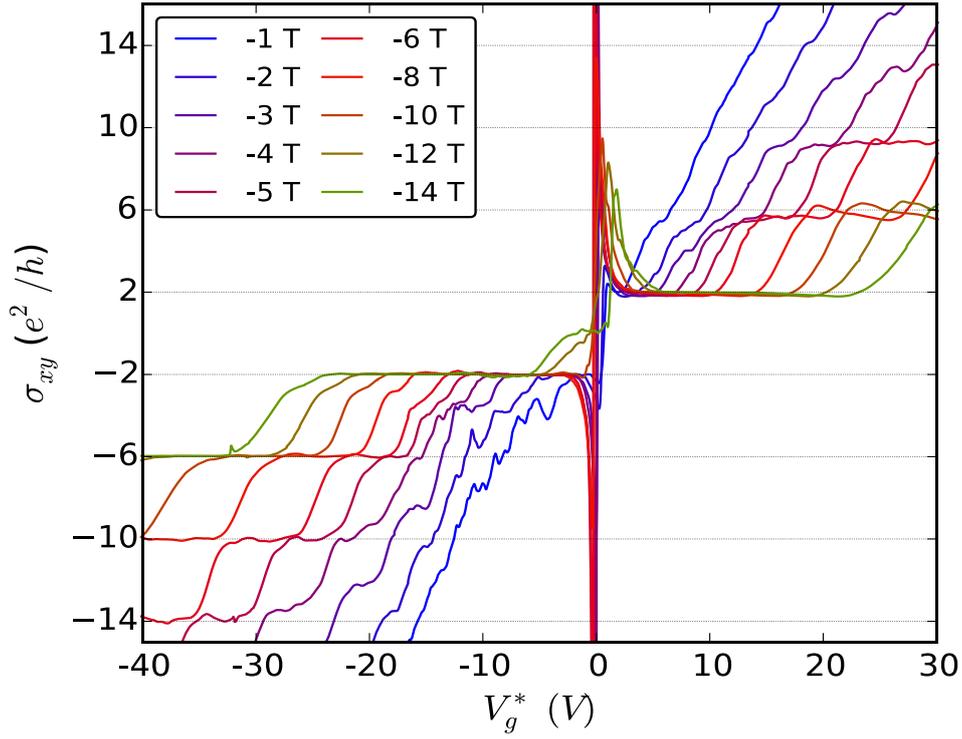

**Figure 5 | Hall conductivity at 1.4 K**. The Hall conductivity $\sigma_{xy}$ ($\sigma_{xy} = \rho_{xy}/(\rho_{xx}^2 + \rho_{xy}^2)$) as a function of $V_g^*$ at different negative magnetic fields. The unconventional sharp conductivity peaks around the Dirac point arise from the longitudinal resistivity dips at the Dirac point. The deviation from strictly quantized conductivity in the electron branch is attributed to the scattering between the quantum Hall states and the conductive $Bi_2Se_3$ channel.

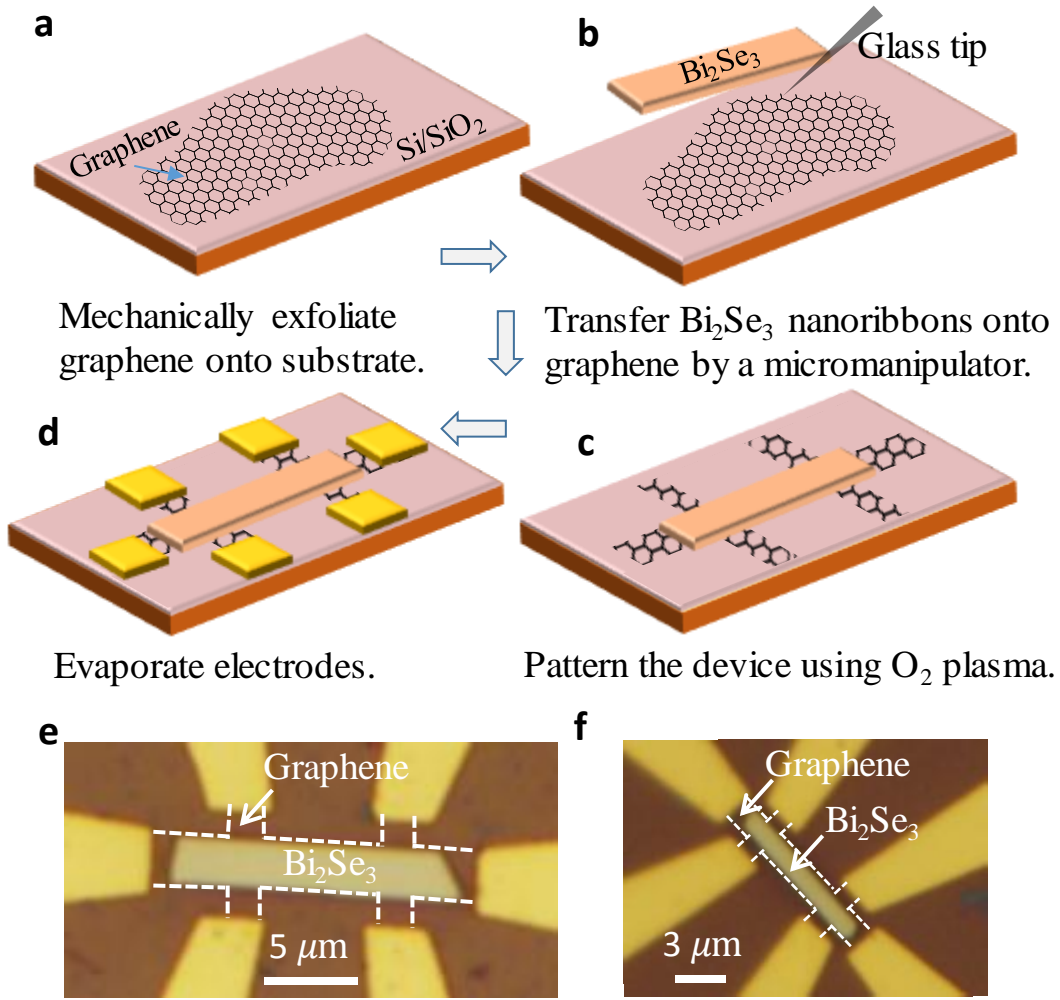

**Supplementary Figure 1: Fabrication of graphene-Bi$_2$Se$_3$ hybrid devices**. The sketches (**a-d**) illustrate the fabrication processes of the graphene-Bi$_2$Se$_3$ heterostructures. (**e,f**) are the optical images of two typical devices. The white dashed lines mark the border of the underlying graphene. The electrodes are only contacted with graphene.

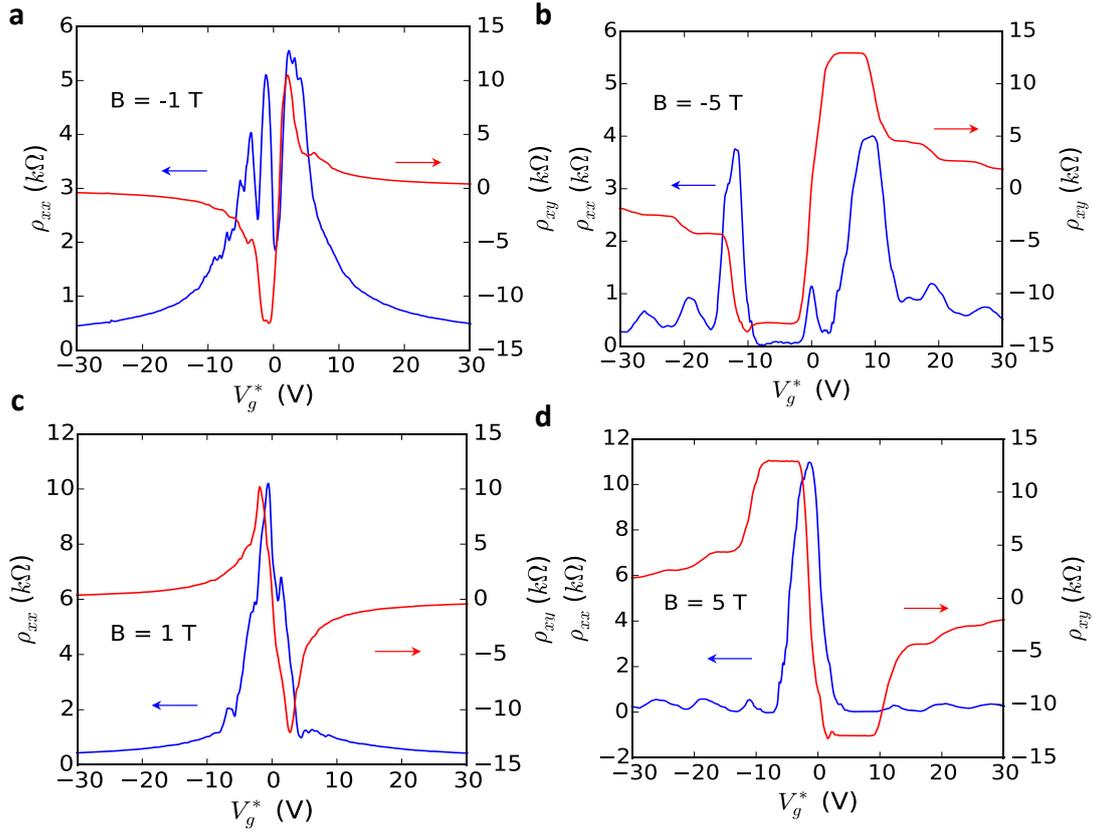

**Supplementary Figure 2: The $\rho_{xx}(V_g^*)$ and $\rho_{xy}(V_g^*)$ curves under both positive and negative magnetic fields for comparison.** The longitudinal resistivity $\rho_{xx}$ (blue curve) and Hall resistivity $\rho_{xy}$ (red curve) as a function of gate voltage $V_g^*$ under various magnetic fields $B$ at 1.4 K. The comparisons between (**a,b**) $\rho_{xx}(V_g^*, B<0)$ and (**c,d**) $\rho_{xx}(V_g^*, B>0)$ directly show the unusual graphene-topological insulator coupling effect. $V_g^*$ is defined as $V_g$ - $V_D$, where $V_D$ is the Dirac point.

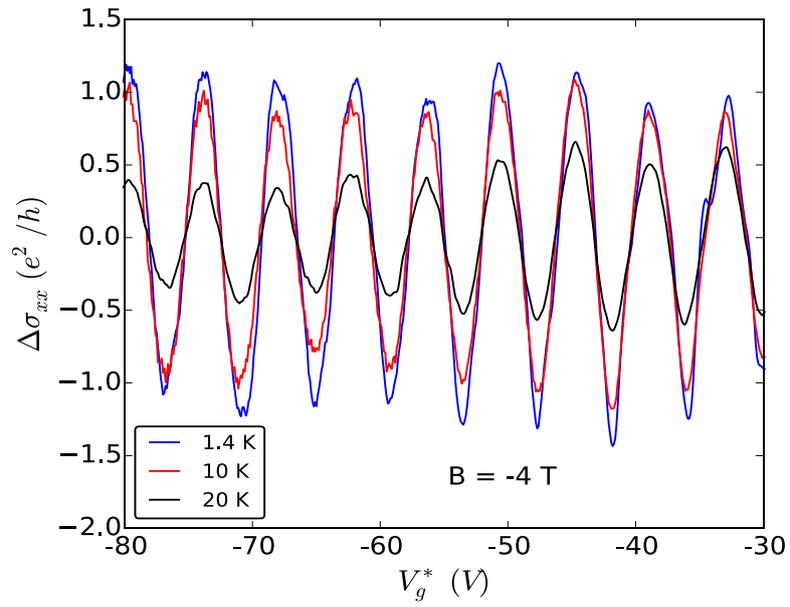

**Supplementary Figure 3: Quantum oscillations in hybrid graphene-topological insulator devices**. The Landau level-induced quantum oscillations under $B = -4$ T are identified.

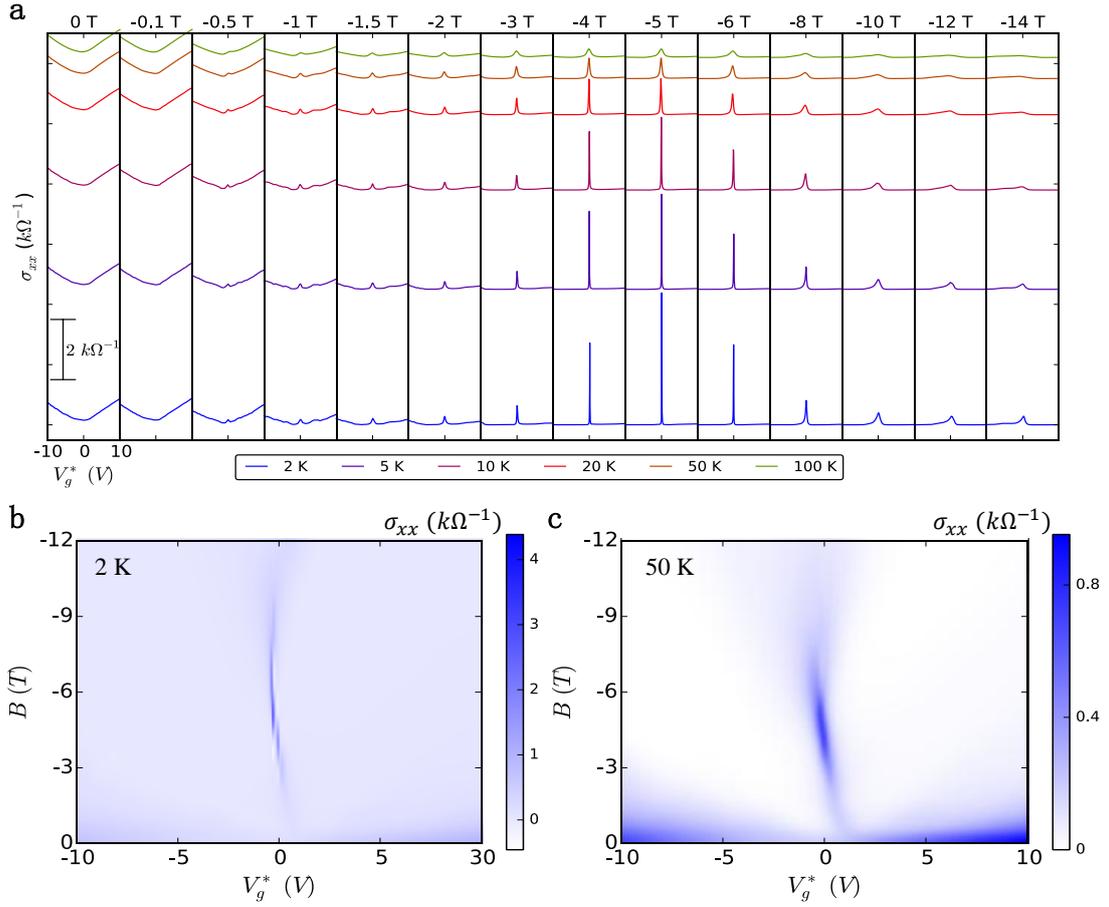

**Supplementary Figure 4: Conductivity evolution near the Dirac point**. (**a**) The evolution of $V_g^*$ dependence of the conductivity $\sigma_{xx}$ ($\sigma_{xx} = \rho_{xx}/(\rho_{xx}^2 + \rho_{xy}^2)$) at various temperatures and magnetic fields. The curves at different temperatures are shifted for clarity. (**b, c**) $\sigma_{xx}$ versus B and $V_g^*$ at 2 K and 50 K, respectively. $V_g^* = V_g - V_D$, where $V_D$ is the Dirac point in $V_g$ under $B = 0$. The position of Dirac point slightly varies under different magnetic fields. The conductivity peaks at the Dirac point under intermediate magnetic fields are very unusual and impressive, which can be ascribed to the magnetic field enhanced coupling between graphene and topological insulator $Bi_2Se_3$. Moreover, the thermal vibration will broaden the peaks and reduce the magnitude.

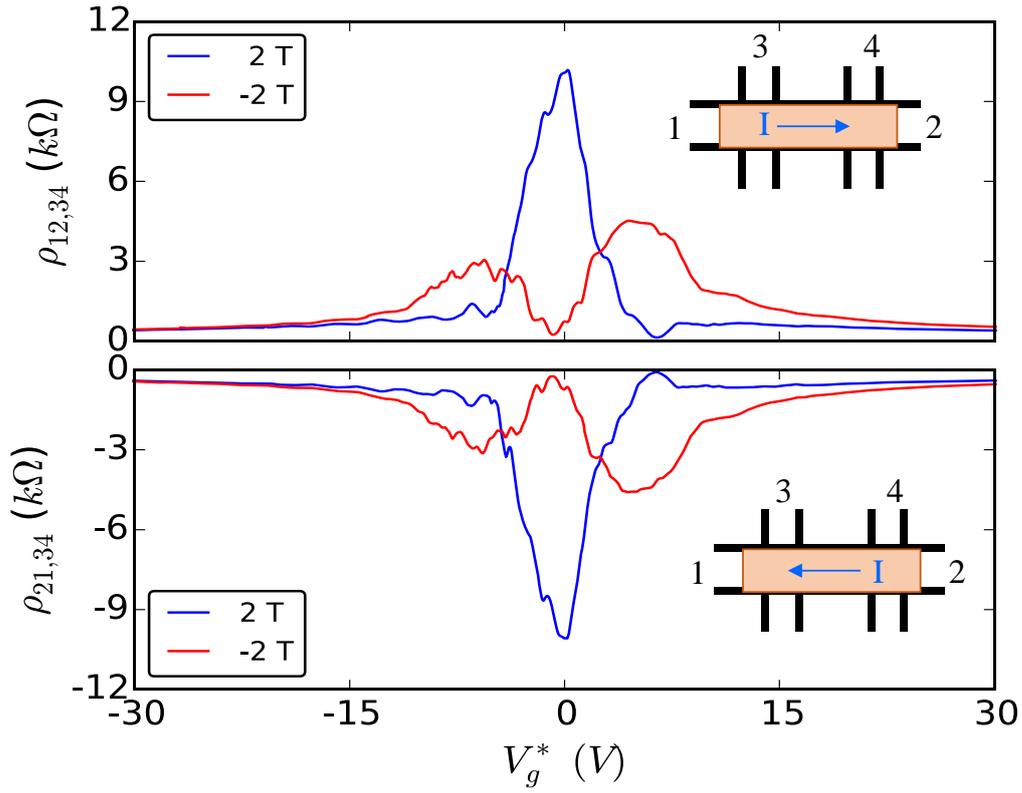

**Supplementary Figure 5: Longitudinal resistivity upon different directions of magnetic field and bias current**. The gate voltage dependence of longitudinal resistivity under reversed direction of magnetic field (B = 2 and -2 T) and current (I = 0.1 and -0.1 μA) at 5 K excludes the Hall origin for the unconventional magnetotransport behaviors. The $\rho_{mn,kl}$ is defined as $\frac{W}{L}\frac{V_{kl}}{I_{mn}}$, where $W$ is the width and $L$ is the length of the Hall bar.

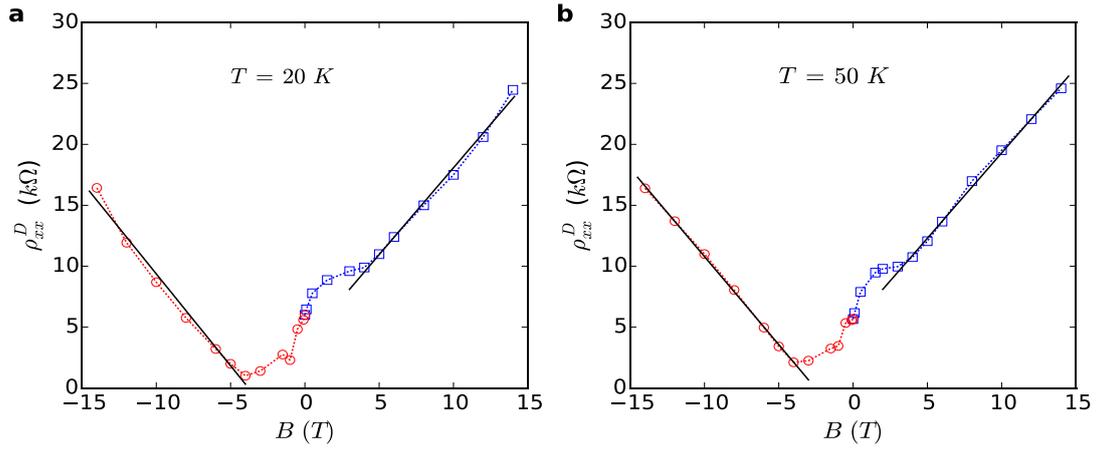

**Supplementary Figure 6: Magnetoresistivity at the Dirac point.** The longitudinal resistivity at the Dirac point $\rho_{xx}^D$ versus magnetic field $B$ at (**a**) 20 K and (**b**) 50 K. The solid lines show the linear fitting results under high magnetic fields. The fitting slopes are 1.43 and -1.51 kΩ/T at 20 K, and 1.4 and -1.44 kΩ/T at 50 K.

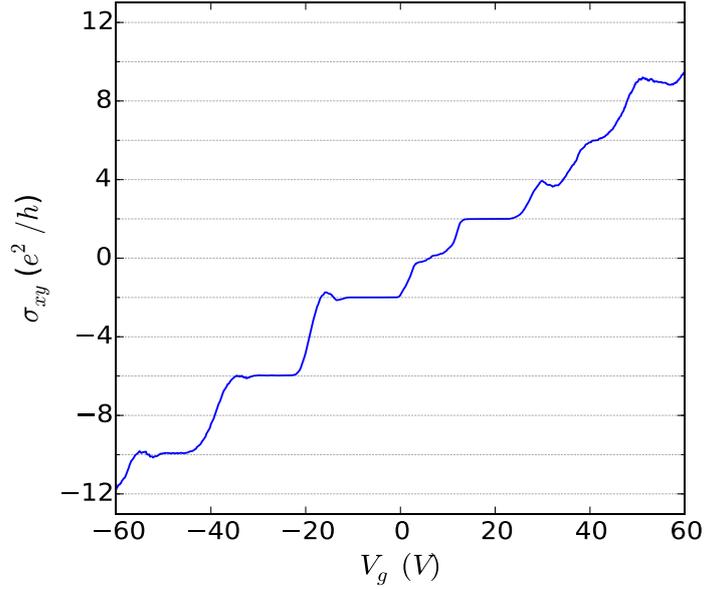

**Supplementary Figure 7: Hall conductivity of another hybrid device with geometry shown in Supplementary Fig. 1f.** The Hall conductivity $\sigma_{xy}$ ($\sigma_{xy} = \rho_{xy}/(\rho_{xx}^2 + \rho_{xy}^2)$) as a function of back gate $V_g$ at 1.4 K and under 14 T measured from another similar graphene-topological insulator hybrid device. The Dirac point locates at ~7 V. In the electron branch, the deviation from strictly quantized Hall conductivity is attributed to the scattering between the quantum Hall edge states and the conducting $Bi_2Se_3$ channels.

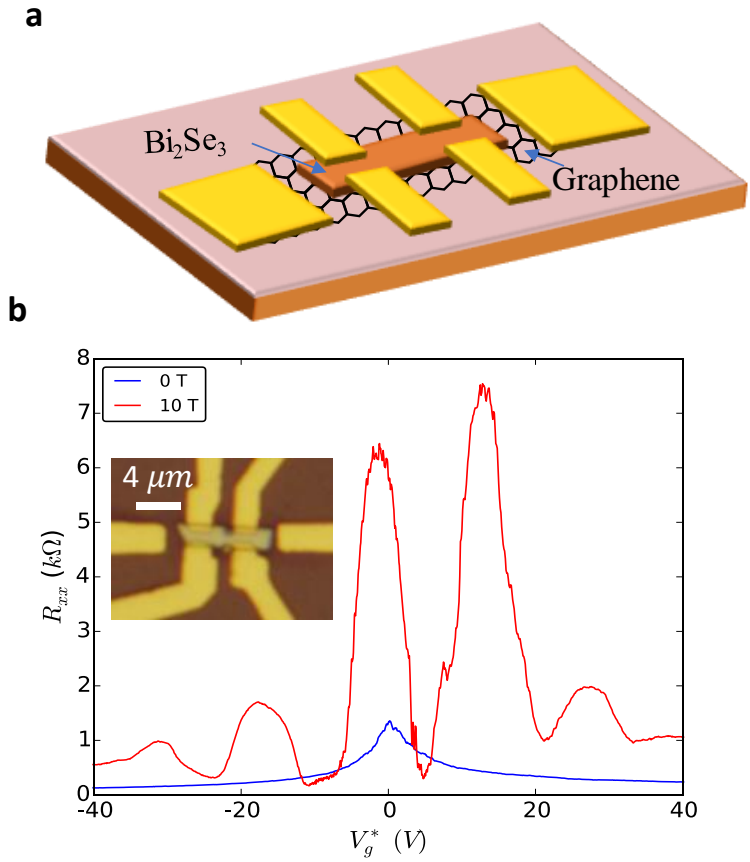

**Supplementary Figure 8: Transport in graphene-Bi$_2$Se$_3$ hybrid device with shared voltage probes at 1.4 K**. (**a**) the schematic diagram of the device. The voltage probes of the Hall bar in graphene-Bi$_2$Se$_3$ heterostructures are contacted with both graphene and Bi$_2$Se$_3$. (**b**) The longitudinal resistance $R_{xx}$ versus gate voltage $V_g^*$. The inset in (**b**) is the optical image of the device. The scattering between quantum Hall sates is significantly increased in the electron branch.

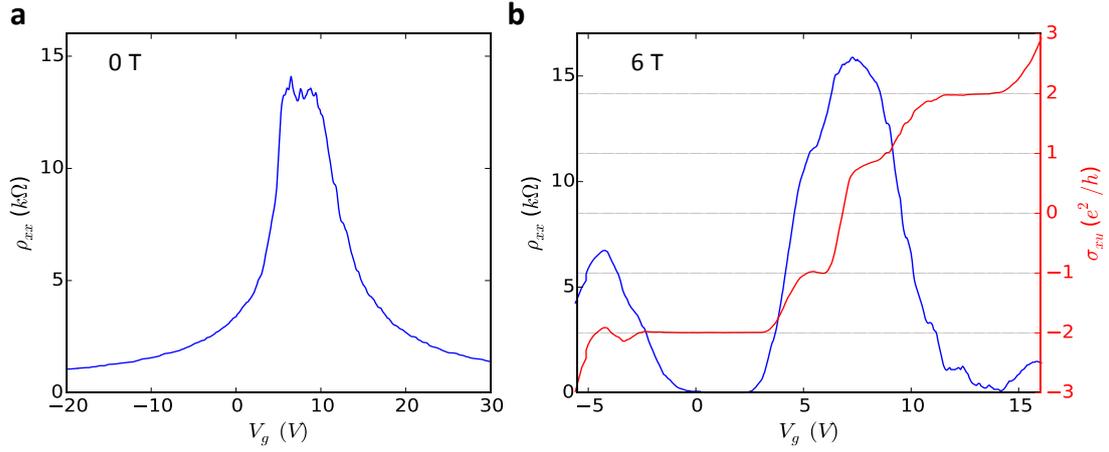

**Supplementary Figure 9: Transport of another hybrid device with geometry shown in Supplementary Fig. 1f**. (**a**) Longitudinal resistivity $\rho_{xx}$ versus back gate voltage $V_g$ without magnetic field. (**b**) $\rho_{xx}$ and Hall conductivity $\sigma_{xy}$ as a function of $V_g$ under 6 Tesla. The graphene-Bi$_2$Se$_3$ hybrid sample was measured at 1.4 K. The Dirac point of this sample locates at ~7 V. The developing $\pm e^2/h$ Hall conductivity plateaus are ascribed to the partially lifted degeneracy of graphene, which is consistent with the theoretical predication of the band structures shown in Fig. 4.